\documentclass[aip,pop,reprint,amsmath,amssymb]{revtex4-1}

\usepackage{graphicx}
\newcommand{\MJ}{Maxwell--J\"{u}ttner }

\def\pop{{\itshape Phys. Plasmas} }

\begin{document}

\title{Modifications to Swisdak (2013)'s rejection sampling algorithm for a \MJ distribution in particle simulations}

\author{Seiji Zenitani}
\affiliation{Space Research Institute, Austrian Academy of Sciences, 8042 Graz, Austria}
\email{seiji.zenitani@oeaw.ac.at}

\begin{abstract}
Modifications to Swisdak [Phys. Plasmas {\bf 20}, 062110 (2013)]'s rejection sampling algorithm for drawing a \MJ distribution in particle simulations are presented. Handy approximations for $e$-folding points and a linear slope in the envelope function are proposed, to make the algorithm self-contained and more efficient.
\end{abstract}

\maketitle


In kinetic plasma simulations
such as particle-in-cell (PIC) and Monte Carlo simulations,
it is often necessary to initialize particle velocities
that follows a certain velocity distribution,
by using random numbers (random variates).
A relativistic Maxwell distribution,
also known as a \MJ distribution,\citep{jut11} is
one of the most important velocity distributions,
in particular in high-energy astrophysics and in laser physics.
Owing to its importance in modeling,
Monte Carlo algorithms for drawing a \MJ distribution
have been developed over many years \citep{swisdak13,zeni22}
(see Ref.~\onlinecite{zeni22} and references therein). 

In the article by \citet{swisdak13},
the author proposed an acceptance-rejection algorithm for a \MJ distribution.
\citet{devroye86}'s piecewise method for log-concave distributions was applied. 
The algorithm is simple and efficient, as will be shown in this paper. 
Technically, it requires an external solver to find $e$-folding points,
where the density is $1/e$ ($= 1/2.71828\dots$) of the maximum. 
This may not be ideal in some applications ---
for example, when the plasma temperature varies from grid cell to cell,
it is necessary to repeatedly call the root finder,
and the user may desire a simpler option. 
In this research note, we propose two modifications to
\citet{swisdak13}'s sampling method.
We propose handy approximations for the $e$-folding points and
a linear slope in the envelope function to improve the acceptance efficiency.

\begin{figure}[htbp]
\includegraphics[width={.9\columnwidth}]{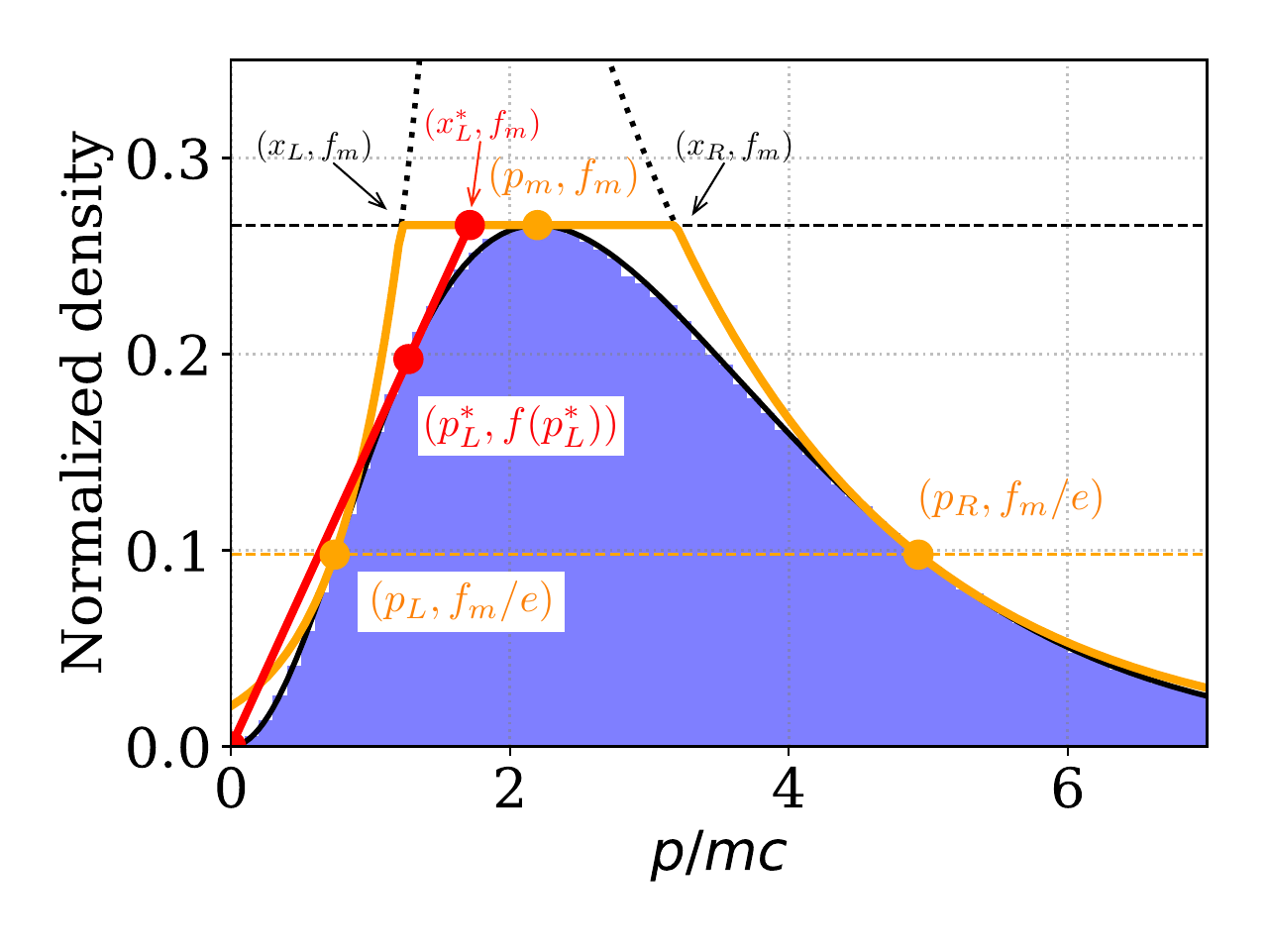}
\caption{The distribution function (black) of a \MJ distribution with $t=1$.
The envelope function (orange) for the rejection method,
the modified envelope (red), and Monte Carlo results (blue histogram) are presented.
\label{fig:swisdak1}}
\end{figure}

First we recap Swisdak's application of the Devroye method.
We limit our attention to an isotropic stationary population.
In the spherical coordinates, the \MJ distribution is given by
\begin{align}
f(p)
=
A p^2
\exp \Big(-\frac{\sqrt{1+p^2} }{t} \Big)
\label{eq:MJ}
\end{align}
where $A$ is the normalization constant,
$p=m\gamma v$ is the momentum, $v$ the velocity,
$\gamma = [1-(v/c)^2]^{-1/2} = [1+(p/mc)^2]^{1/2}$ the Lorentz factor,
and $t=T/mc^2$ the temperature.
For simplicity, we set $m=c=1$. 
A detail form of the normalization constant is
found in many literature,\citep{zeni22}
but this article does not rely on it. 
The black curve in Fig.~\ref{fig:swisdak1} shows
the distribution function with $t=1$.
It has a maximum
$f_m = f(p_m)$
at $p_m = [ 2 t \left( t + \sqrt{1+t^2} \right) ]^{1/2}$. 
The rejection algorithm uses a piecewise envelope function $F(p)$  of two exponential tails and a flat line,
as shown in orange in Fig.~\ref{fig:swisdak1}.
\begin{align}
F(p)
&=
\left\{
\begin{array}{ll}
f(p_{L}) \exp\left(\dfrac{p-p_L}{\lambda_L}\right)
~~
& (p \le x_L)
\\ 
f_m
~~
& (x_L < p \le x_R)
\\ 
f(p_{R}) \exp\left(-\dfrac{p-p_R}{\lambda_R}\right)
& (x_R < p)
\end{array}
\right.
\label{eq:envelope1}
\end{align}
Here, the exponential tails
touch the distribution function at $p=p_L$ and $p_R$,
$\lambda_X=|f(p_X)/f'(p_X)|$ is the scale length
at $p=p_X$ ($X$=$L, R$), and
the two switching points are located at
\begin{align}
x_L = p_L + \lambda_L \log \frac{f_m}{f(p_L)}
,~~
x_R = p_R - \lambda_R \log \frac{f_m}{f(p_R)}
\label{eq:xR}
\end{align}
The densities of the three parts of Eq.~\eqref{eq:envelope1} are $f_m {\lambda_L}, f_m (x_R - x_L )$, and $f_m {\lambda_R}$.
Based on them, the rejection algorithm was constructed accordingly.\citep{devroye86,swisdak13}

Importantly, $p_L$ and $p_R$ are given as inputs.
It was shown in \citet{devroye86} that the acceptance rate is highest when
\begin{align}
f(p_L) = f(p_R) = \frac{f_m}{e}
\label{eq:swisdak_e}
\end{align}
\citet{swisdak13} numerically obtained
such solutions by using a root finder.
This choice makes several terms simpler,
as presented in Ref.~\onlinecite{swisdak13}.


\begin{figure}[tbhp]
\includegraphics[width={\columnwidth}]{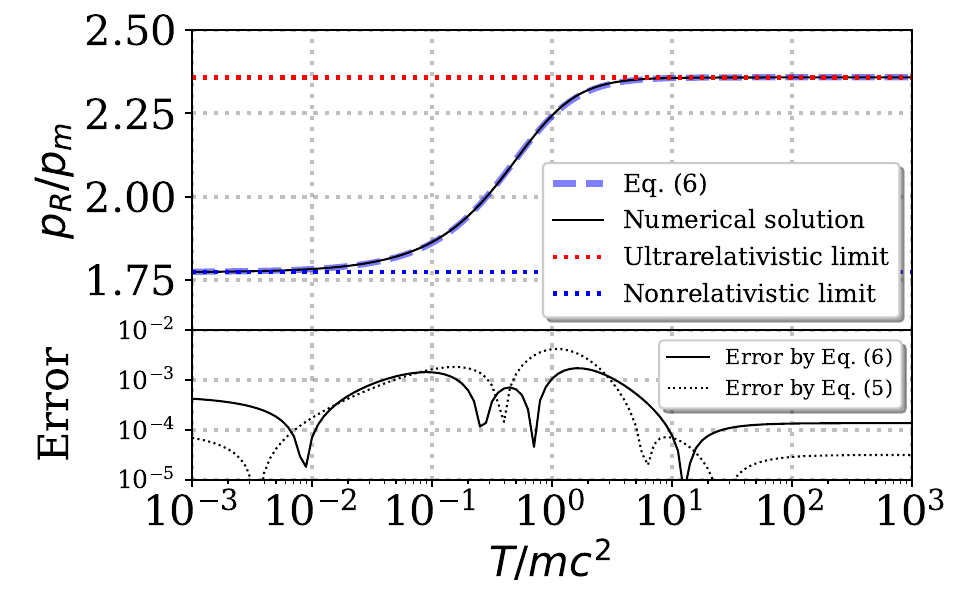}
\caption{
Right $e$-folding position $p_R$, as a function of $t$.
The bottom plot shows the relative errors of
the two approximations on a log scale. 
\label{fig:approx}}
\end{figure}

Next we present our modification.
We propose the following approximations for the $e$-folding points,
\begin{align}
p_L &\approx \left( 0.3017 + \dfrac{0.386}{4+3 t+4 t^2} \right) p_m
\label{eq:formula1}\\
p_R &\approx \left( 2.358 - \dfrac{1.168}{2+3 t+5 t^2} \right) p_m
\label{eq:formula2}
\end{align}
Below we explain how we derived them. 

In the nonrelativistic limit of $t \rightarrow 0$,
the distribution function and the mode are reduced to
\begin{align}
f(p) = A p^2 \exp\left( -\dfrac{1}{t} -\dfrac{p^2}{2t} \right),~~
p_m = \sqrt{2t}
\label{eq:nonrela}
\end{align}
By solving Eqs.~\eqref{eq:swisdak_e} and \eqref{eq:nonrela}, we find
${p_L}/{p_m} = \sqrt{ -W_{0}\left(-{e^{-2}}\right)} \approx 0.3977$ and
${p_R}/{p_m} = \sqrt{ -W_{-1}\left(-{e^{-2}}\right)} \approx 1.774$.
Here, $W_{0}(x)$ and $W_{-1}(x)$ are
the upper and lower branches of the Lambert W function.
In the ultrarelativistic limit of $t \rightarrow \infty$,
they are asymptotic to
\begin{align}
f(p) = A p^2 \exp\left(-\frac{p}{t} \right),~~
p_m = {2t}
\label{eq:ultrarela}
\end{align}
Equations \eqref{eq:swisdak_e} and \eqref{eq:ultrarela} give
$p_L/{p_m} = -W_{0}\left(-{e^{-1.5}}\right) \approx 0.3017$ and
$p_R/{p_m} = -W_{-1}\left(-{e^{-1.5}}\right) \approx 2.358$. 

We assume that $p_X/p_m$ ($X=L, R$) is approximated by
a rational function of $t$:
\begin{align}
\dfrac{p_X}{p_m} = \frac{a+bt+ct^2}{1+qt+rt^2}
\end{align}
where $a,b,c,q,$ and $r$ are positive parameters.
Its derivative is
\begin{align}
\left( \dfrac{p_X}{p_m} \right)'
&= \frac{ b - qa + 2(c-ra)t + (cq-br)t^2}{(1+qt+rt^2)^2}
\end{align}
We assume that $p_X/p_m$ changes slowly.
By setting  $b=cq/r$, we eliminate the second order term in the numerator.
Then $p_X/p_m$ monotonically increases/decreases
from $a$ at $t=0$ to $c/r$ in the $t\rightarrow \infty$ limit,
\begin{align}
\dfrac{p_X}{p_m}
&= \frac{c}{r} + \frac{a-(c/r)}{1+qt+rt^2}
.
\end{align}
The two limits $a$ and $c/r$ are set to the asymptotic values, discussed earlier.
Once they are given, we search for $q$ and $r$.
We assume that $q$ and $r$ are fractions, 
to keep the final equations simple. 
Considering these issues,
we obtain Eqs.~\eqref{eq:formula1} and \eqref{eq:formula2}
by trial and error.

The top portion of Fig.~\ref{fig:approx} compares
Eq.~\eqref{eq:formula2} and the numerical solution of
the right $e$-folding point $p_R$,
as a function of $t$. 
The bottom portion shows
the relative error between Eq.~\eqref{eq:formula2} and $p_R$,
and also
the relative error between Eq.~\eqref{eq:formula1} and $p_L$. 
Note that the numerical solutions are obtained by a root finder
(the {\tt scipy.optimize.fsolve} function in Python),
whose tolerance is $\approx 1.5 \times 10^{-8}$. 
These plots show that
Eqs.~\eqref{eq:formula2} and \eqref{eq:formula1}
are very good approximations. 
Meanwhile, since these approximations do not guarantee Eq.~\eqref{eq:swisdak_e},
the logarithmic terms in Eq.~\eqref{eq:xR} should be retained, but \citet{devroye86}'s original procedure works. 

\begin{table}[tpb]
\centering
\caption{A modified algorithm with a linear slope
\label{table:swisdak3}}
\begin{tabular}{l}
\hline
require: $t$\\
$f(x) \equiv p^2 \exp(-\sqrt{1+p^2}/t)$ \\
\hline
$p_m^2 \leftarrow 2 t \left( t + \sqrt{1+t^2} \right)$,~
$f_m \leftarrow f(p_m)$ \\
$(p^{*}_L)^2 \leftarrow \frac{1}{2}\left( { t^2 + t\sqrt{4+t^2} } \right)$,~
$x^{*}_L \leftarrow \dfrac{f_m}{f(p^{*}_L)}p^{*}_L$ \\
$p_R \leftarrow \left( 2.358 - \dfrac{1.168}{2+3 t+5 t^2} \right) p_m$\\
$\lambda_R \leftarrow -\dfrac{f(p_R)}{f'(p_R)}$,~
$x_R \leftarrow p_R + \lambda_R \log \dfrac{f(p_R)}{f_m}$
\\
$S \leftarrow x_R - \frac{1}{2} x^{*}_L + \lambda_R $\\
$q_L \leftarrow \dfrac{x^{*}_L}{2S}$,~
$q_R \leftarrow \dfrac{\lambda_R}{S}$,~
$q_C \leftarrow 1 - q_L - q_R$\\
{\bf repeat}\\
~~~~generate $X_1, X_2 \sim U(0, 1)$ \\
~~~~{\bf if}~$X_1 < q_L$ {\bf then} ~~~// Left slope \\
~~~~~~~~$p \leftarrow x^{*}_L\sqrt{ X_1/q_L }$ \\
~~~~~~~~{\bf if}~$p X_2 \le f(p)x^{*}_L/f_m$ {\bf break} \\
~~~~{\bf else if}~$X_1 \le q_L+q_C$ {\bf then}~~~~ // Central box \\
~~~~~~~~$p \leftarrow x^{*}_L + (x_R-x^{*}_L) ((X_1-q_L)/q_C)$ \\
~~~~~~~~{\bf if}~$X_2 \le f(p)/f_m$ {\bf break} \\
~~~~{\bf else} ~~~~~~~~~~~~~~~~~ // Right tail \\
~~~~~~~~$U \leftarrow (X_1-q_C-q_L)/{q_R} $ \\
~~~~~~~~$p \leftarrow x_R - \lambda_R \log U $ \\
~~~~~~~~{\bf if}~$U X_2 \le f(p)/f_m$ {\bf break} \\
~~~~{\bf endif}\\
{\bf end repeat}
\\
generate $X_3, X_4 \sim U(0, 1)$ \\
$p_x \leftarrow p ~ ( 2 X_3 - 1 )$ \\
$p_y \leftarrow 2 p \sqrt{ X_3 (1-X_3) } \cos(2\pi X_4)$ \\
$p_z \leftarrow 2 p \sqrt{ X_3 (1-X_3) } \sin(2\pi X_4)$ \\
\hline
\end{tabular}
\end{table}

\begin{figure}[tbhp]
\includegraphics[width={\columnwidth}]{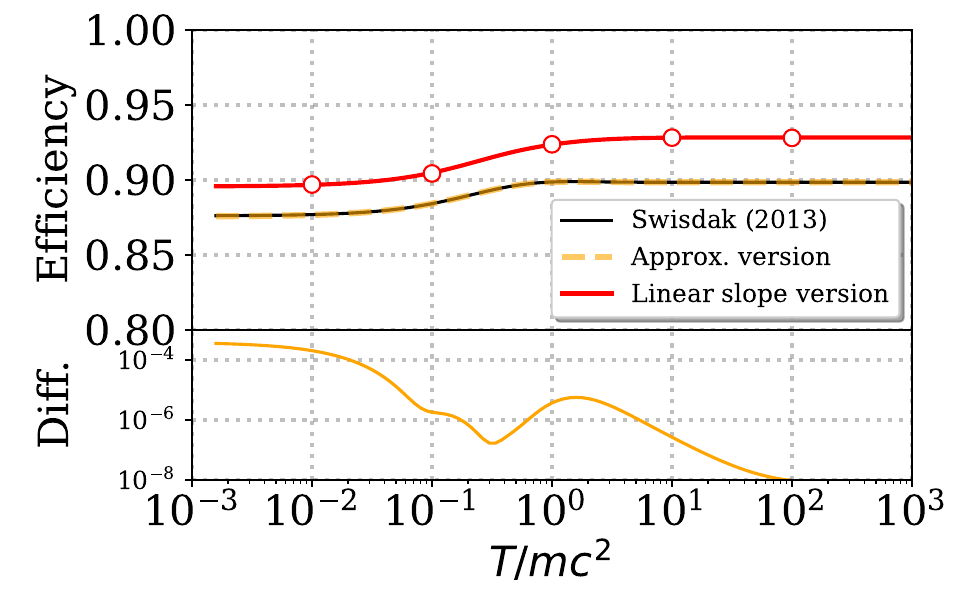}
\caption{
Acceptance efficiencies of
the original Swisdak method (black),
the approximate version (orange), and
the final version with the linear slope (Table~\ref{table:swisdak3}; red).
The red circles indicate Monte Carlo results.
The bottom panel shows the difference
between the original method and the approximate method.
\label{fig:swisdak_eff}}
\end{figure}

Next, we present our second modification.
As shown by the red line in Fig.~\ref{fig:swisdak1},
we replace the left tail by a linear slope in the envelope function ,
\begin{align}
F^{*}(p)
&=
\left\{
\begin{array}{ll}
f_m \left( \dfrac{p}{x^{*}_L} \right)
~~
& (p \le x^{*}_L)
\\
f_m
~~
& (x^{*}_L < p \le x_R)
\\
f(p_{R}) \exp\left(-\dfrac{p-p_R}{\lambda_R}\right)
& (x_R < p)
\end{array}
\right.
\label{eq:envelope2}
\end{align}
The slope touches the distribution function 
at $(p^{*}_L)^2 = \frac{1}{2} \left({ t^2 + t \sqrt{ 4 + t^2  } }\right)$,
obtained from
$\frac{\partial}{\partial p} \left( {f(p)}/{p} \right) = 0$. 
The slope meets the maximum line at
$x^{*}_L = \{{f_m}/{f(p^{*}_L)}\} p^{*}_L$.
Its relative position $x^{*}_L/p_m$ monotonically changes
from $(2/e)^{1/2} \approx 0.858$ ($t \rightarrow 0$)
to $(2/e) \approx 0.738$ ($t \rightarrow \infty$).
It is easy to generate a triangle-shaped distribution for
the left part of Eq.~\eqref{eq:envelope2}, by using a random variate. 
The partial densities of Eq.~\eqref{eq:envelope2} are
$\frac{1}{2}f_m {x^{*}_L}$, $f_m (x_R - x^{*}_L )$, and $f_m {\lambda_R}$, and then the rejection scheme can be constructed, accordingly. 
A formal algorithm is presented in Table \ref{table:swisdak3}.
For completeness, a procedure to scatter $p$
in three dimensions is added to the last part,
so that the algorithm is self-contained.
In practice, we can use Eq.~\eqref{eq:MJ} with $A=1$,
because the algorithm is independent of $A$.


We have generated a \MJ distribution of $10^6$ particles with $t=1$,
using the final algorithm with the linear slope.
The results are shown by the blue histograms in Fig.~\ref{fig:swisdak1},
in agreement with the distribution (the black curve).
This demonstrates that the proposed methods are ready to use in PIC and Monte Carlo simulations.
The top portion of Fig.~\ref{fig:swisdak_eff} compares
theoretical acceptance rates of the original Swisdak method, 
the approximate version with Eqs.~\eqref{eq:formula1} and \eqref{eq:formula2},
and the linear-slope version in Table~\ref{table:swisdak3}.
As already reported,\citep{swisdak13}
the Swisdak method gives $88$--$90\%$. 
The approximate version gives very similar results,
thanks to the good approximations of the $e$-folding points.
This is evident in the (absolute) error
in the bottom portion of Fig.~\ref{fig:swisdak_eff}. 
The acceptance rate of the linear-slope version is
\begin{align}
\dfrac{1}{f_m(x_R - \frac{1}{2}x^{*}_L + \lambda_R)} = 90{\rm -}93\%
.
\label{eq:eff}
\end{align}
This prediction is confirmed by Monte Carlo tests with $10^6$ particles,
shown in open circles.
The efficiency is improved by a few percent,
because the linear slope bounds the distribution better. 
These three versions are as efficient as today's leading methods.\citep{zeni22}
By comparing Fig.~\ref{fig:swisdak_eff} with Fig. 9 in Ref.~\onlinecite{zeni22},
the reader will see that they are very competitive
in acceptance efficiency.

The proposed modifications are moderate, and would be rewarded by the numerical cost of the root finder and by the better acceptance efficiency.

\section*{Acknowledgements}
The author acknowledges Marc Swisdak for discussion.

\section*{Conflict of Interest}
The author has no conflicts to disclose.

\section*{Data Availability}
A Jupyter notebook is available at \url{https://doi.org/10.5281/zenodo.13274577}.

\end{document}